# Impact of interfaces on photoluminescence efficiency of high indium content InGaN quantum wells

P. Wolny[1], H. Turski[1], G. Muziol[1], M. Sawicka[1], J. Smalc-Koziorowska[1], J. Moneta[1], A. Feduniewicz-Żmuda[1], S. Grzanka[1], C. Skierbiszewski[1]

[1]Institute of High Pressure Physics, Polish Academy of Sciences, Warsaw, 01-142, Poland

**Abstract**

InGaN-based light emitting diodes (LEDs) are known to suffer from low electron and hole wavefunction overlap due to high piezoelectric field. Staggered InGaN quantum wells (QWs) have been proposed to increase the wavefunction overlap and improve the efficiency of LEDs especially for long wavelength emitters. In this work we evidence that the growth of staggered QWs has also another beneficial effect as it allows to reduce the formation of defects, responsible for nonradiative Shockley-Read-Hall recombination, at the bottom interface of the QW. Staggered QWs comprised an InGaN layer of an intermediate In content between the barrier and the QW. We show that insertion of such a layer results in a significant increase of the luminescence intensity, even if the calculated wavefunction overlap drops. We study the dependence of the thickness of such an intermediate In content layer on photoluminescence (PL) intensity behavior. Staggered QWs exhibit increased cathodoluminescence (CL) homogeneity that is a fingerprint of lower density of defects, in contrast to standard QWs for which high density of dark spots are observed in QW emission mapping. Transmission electron microscopy of standard QWs revealed formation of basal-plane stacking faults (BSFs) and voids that could have resulted from vacancy aggregation. Stepwise increase of the In content in staggered QWs prevents formation of point defects and results in an increased luminescence efficiency. The In composition difference between the barrier and the well is therefore a key parameter to control the formation of point defects in the high-In content QWs, influencing the luminescence efficiency.

**Introduction**

The fabrication of nitride based emitters still poses a challenge when long emission wavelengths in the green part of the spectrum are considered. There are two main reasons for the existence of the so called "green gap". The first is a consequence of the presence of large electric fields in GaN/InGaN quantum wells (QWs), which separate the free carriers and decrease the wavefunction overlap [1, 2]. Especially for high indium content QWs, the built-in electric field can significantly decrease quantum efficiency of nitride structures, due to the decrease of the electron and hole wavefunction overlap. The second reason is related to more challenging InGaN epitaxy leading to defect formation [3, 4]. The optimization of structural and optical quality of InGaN QWs addressing the thermodynamic and structural limitations, i.e. overcoming high decomposition rate of In-N bonds while preserving high optical quality of high In content InGaN [5-7]. Furthermore, the incorporation limits of indium to the InGaN layers due to the lattice mismatch effects were recently intensively investigated. The interesting results, obtained both experimentally [8, 9] and theoretically [10], point out that InGaN monolayer coherently grown on GaN substrate can have an In content of only 33%. Therefore, in order to obtain a higher In content in the coherently grown InGaN, an increase of the *a* lattice constant of the substrate is needed [11]. Additionally, proper strain engineering is required to control the large lattice mismatch in high indium content QWs to avoid generation of misfit defects [12, 13].

The impact of the growth conditions on the quantum efficiency (QE) is also important for blue LEDs. We would like to point out that an increase of the QE in blue LEDs by introducing the InGaN interlayer before the QWs for MOVPE growth was recently discussed very intensively [14, 15]. According to one of the hypothesis, the QE collapse can been ascribed to the generation of surface defects (vacancies) in GaN at high growth temperatures, which are then trapped by InGaN QWs [14]. The other report points out on the unintentional doping to be the reason of the presence of nonradiative recombination centers [15]. As an explanation, in both cases - the InGaN interlayer is important to bury these defects (vacancies or impurities) before QW is grown.

The concept of staggered QWs has been utilized for green emitters for quite some time already. Originally, it was proposed in order to increase the electron-hole wavefunction overlap [16, 17]. Interestingly, in the light-emitting diode (LED) structures grown by metal-organic vapor phase epitaxy (MOVPE) the enhancement in electroluminescence from LEDs with staggered QWs was larger than that predicted theoretically. However, the authors only speculated about the reason for it mentioning: (1) more pronounced carrier screening that reduces energy band bending and further increases overlap of $\Gamma_{e-hh}$, (2) improved material quality, and (3) better carrier confinement in the staggered InGaN QWs.

In this work we discuss the origin of the increased photoluminescence (PL) of staggered InGaN QWs grown by plasma-assisted molecular beam epitaxy (PAMBE) in respect to their standard counterparts. The behavior of PL is investigated taking into account two effects (1) electron-hole wavefunction overlap and (2) generation of extended and point defects (vacancies) at the interface between the barrier and the QW. Our study presents a comparison of standard and staggered QWs. We show that the insertion of an InGaN layer of an intermediate In content (which we will call a preQW) just before the QW has a beneficial effect on the luminescence efficiency of the staggered QW. The increase in the luminesce intensity observed in staggered QWs can be attributed to reduced density of defects, which is due the smaller difference in composition at InGaN interfaces in comparison with standard QW. We propose that defects which reduce PL intensity are related to the generation of a large number of vacancies when standard, high In content InGaN QW is grown. Additional argument supporting the vacancies being generated at the interface between InGaN layers that greatly differ in In content is the fact that the vacancy aggregation can result in basal-plane stacking faults (BSF) and voids formation [18]. Presence of such defects is confirmed experimentally by transmission electron microscopy (TEM).

**Experimental**

The QW structures studied in this work were grown by PAMBE in a custom-designed Gen20A reactor (Veeco Instruments). Bulk GaN substrates with a threading dislocation density of $1\cdot10^7 cm^{-2}$ were used. Indium-rich growth conditions were employed to ensure the step-flow surface morphology at growth temperature of 650 °C [19]. The presence of indium on the surface after the growth as well as growth temperature were monitored by laser reflectometry [20]. The InGaN growth conditions needed to obtain a specific In content were calibrated separately on another set of samples for which the In content was assessed by X-Ray diffraction. The photoluminescence was measured at room temperature with a continuous work He-Cd laser operating at λ=325nm. The excitation power density was about 50 W/cm$^2$. Conventional, high-resolution TEM, High Angle Annular Dark Field (HAADF) scanning TEM (STEM) and quasi bright field (BF) STEM by using HAADF detector with large camera length (1.8 m) were performed on an FEI TITAN 200 transmission electron microscope operated at 200 keV in cross-section for the selected samples.

Details of all the structures discussed in this work are listed in Table 1. Three series of samples were grown. In each of the series, two types of structures were grown: standard (without preQW) and staggered (with preQW). The first and the second series consisted of double 2.6nm InGaN QWs with an In content of 22% and 24%, respectively. The third series consisted of a single 12nm InGaN QW with In content of 22%. Samples with thin QWs contained two QWs, while samples with wide QWs had only a single QW. Schematic diagrams of the composition profiles in the thin double QWs and wide single QW are presented in Figure 1(a) and Figure 1 (b), respectively. PreQW thickness is denoted as d. Note that for d=0 there is no preQW. The thickness of preQW InGaN layer was varied from 0 to 5 nm in the third series of samples with wide QWs. The In content in InGaN barriers between the QWs and cap layer was the same in all of the samples and equal to 8%. The In content in the preQW layer was 16%. Table 1 presents also the peak maximum position and relative intensity of room temperature PL as well as electron-hole wavefunction overlaps numerically calculated for single QWs for given QW and preQW thickness.

*Table 1. List of the samples studied with the thickness of preQW and QW indicated. The electron-hole wavefunction overlaps numerically calculated for single QWs for given QW and preQW thickness. (\*For wide QWs, the values are given only for the most probable transition). Room-temperature PL maximum wavelength are presented together with relative intensity.*

| Sample | Number of QWs | PreQW (nm) | QW (nm) | $\langle\psi_e|\psi_h\rangle$ at j=100 A/cm$^2$ | Room temperature PL QW peak wavelength (nm) | intensity (% of the reference) |
|---|---|---|---|---|---|---|
| A0 | 2 | 0 | 2.6 | 0.40 | 491 | 22 |
| A1 | 2 | 1 | 2.6 | 0.33 | 488 | 56 |
| B0 | 2 | 0 | 2.6 | 0.38 | 512 | 4 |
| B1 | 2 | 1 | 2.6 | 0.32 | 512 | 13 |
| W0 | 1 | 0 | 12 | 0.45* | 479 | 1 |
| W02 | 1 | 0.2 | 12 | 0.42* | 475 | 74 |
| W1 | 1 | 1 | 12 | 0.35* | 480 | 100 |
| W2 | 1 | 2 | 12 | 0.38* | 480 | 92 |
| W5 | 1 | 5 | 12 | 0.47* | 478 | 92 |

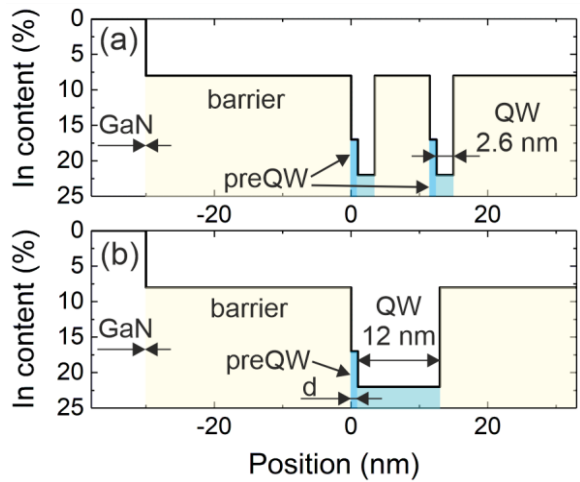

Figure 1. Schematic diagram presenting the studied QW structures that comprise preQW of a thickness d inserted before: (a) 2.6 nm thick double QWs and (b) 12 nm thick single QW.

## Results and discussion

**Photoluminescence studies**

The room-temperature PL spectra of the three sample pairs are shown in Figure 2 (a-c). The first pair A0-A1 constitute of double QW structures that have 22% In in the InGaN QW (2.6 nm thickness) and emit at about 490 nm. Sample A0 has a standard, abrupt composition profile of the QWs (d = 0 nm) while in sample A1, each QW is preceded by the 1-nm thick preQW InGaN layer with In composition of 16%. A similar sample pair is represented by B0 and B1 (QW thickness 2.6 nm), but here the In content of the QWs is 24% that results in longer emission wavelength with the peak maximum at 512 nm. The last sample pair is W0 and W1 that represent the results for single, 12-nm-thick QWs of 22% In, emitting at 480 nm. The PL spectrum of each sample was measured under identical conditions. For all three pairs of samples we observe higher emission intensity for the staggered structure regardless of the QW In content or width.

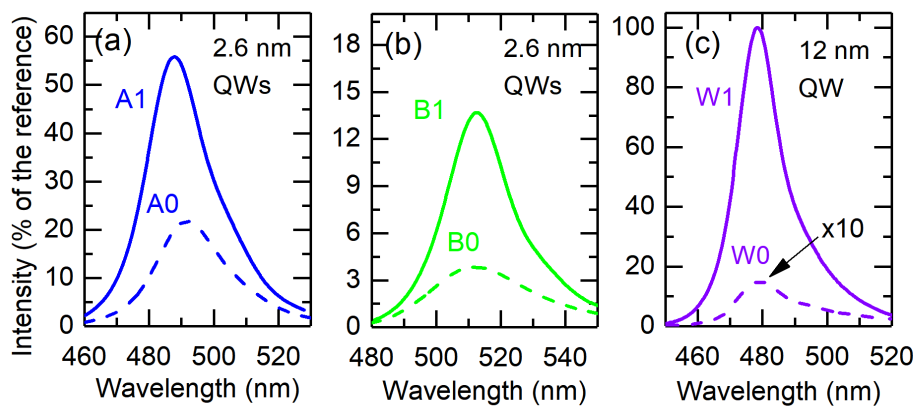

Figure 2. Comparison of RT PL of staggered and standard QWs: double QW structures with 2.6 nm thick QWs having (a) 22% In and (b) 24% In in the QWs and (c) 12 nm thick single QW of 22% In content. PL of the standard structures and staggered structures with preQW thickness of 1 nm are presented in dashed and solid line, respectively.

The impact of the insertion of the preQW on the electron-hole (e-h) wavefunction overlaps, $\langle\psi_e|\psi_h\rangle$, was verified theoretically. Energy band structures and electron and hole wavefunctions were calculated for all the structures under study using SiLENSe 6.4 [21]. Single QWs 2.6 nm and 12 nm thick, with or without a preQW were considered. It is worth to mention that in case of 12 nm thick QW the wavefunction overlap between the ground states is negligible, however there exists a transition path through excited states, which probability can be exceptionally high [22]. The applicability of wide QWs was confirmed by implementing them into laser diode structures [22, 23]. In case of 12 nm wide QWs there are several transitions which can participate in carrier recombination. However, for the purpose of this work, we restrict ourselves to qualitative comparison, presenting the only the highest $\langle\psi_e|\psi_h\rangle$ value for the respective structures in Table 1. Furthermore, despite the e-h wavefunction overlaps were calculated for single QW, the results still well illustrate the experimentally studied structures with two QWs (A0, A1 and B0, B1) that are far apart. One can see that $\langle\psi_e|\psi_h\rangle$ in thin QWs with a preQW are in fact lower than the standard QWs. This is a consequence of the fact that the total QW thickness increased significantly after insertion of the preQW layer. Such effect is not so well visible for wide QW, where $\langle\psi_e|\psi_h\rangle$ is high for the standard and for staggered structures. In case of the third series with 12 nm wide QWs we observe that the impact of the preQW on the wavefunction overlap is negligible.

Two selected cases that illustrate the A1 and W1 samples are presented in Figure 3. Energy band structures and respective $\langle\psi_e|\psi_h\rangle$ in thin and thick QWs with a 1-nm-thick preQW without and under excitation (j=100A/cm$^2$) are plotted. One can see that observed emission from wide QWs will come from the excited states e2-h2 and e2-h3 (see Figure 3b), while the wavefunction overlap between the ground states is negligible.

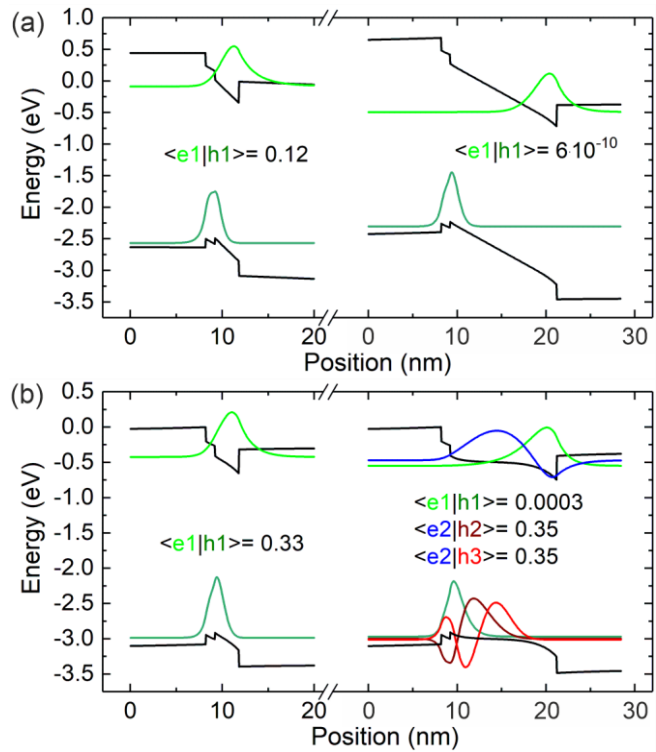

*Figure 3. Calculated band diagrams for single thin (left) and wide (right) InGaN QWs with 1-nm-thick preQW: (a) without and (b) under excitation. The e-h overlaps are denoted. Note that under excitation, the e2-h2 and e2-h3 transitions dominate in wide QWs.*

Impact of the width of the preQW layer was investigated in the series W0-W5 with a variable preQW thickness, d, from 0 to 5 nm. The 12-nm-thick QWs are intentionally used and compared in this experiment in order to analyze and contrast the results with the 2.6-nm-thick QWs. The summary of the PL intensity with respect to preQW thickness is presented in Figure 4 while the details of the peak wavelength position are listed in Table 1. Surprisingly, for standard wide QWs (sample W0) we observe the lowest PL intensity. The situation is dramatically changed for staggered thick QWs (W02-W5). The PL intensity measured for the structures with preQW is significantly higher than for the standard one. The highest PL intensity among all is recorded for the staggered QW structure W1 and it is taken as the reference intensity for all other samples. What is interesting, the growth of one monolayer of an intermediate In content preQW results in a dramatic increase of PL, and an increase in preQW thickness does not bring further significant improvement in intensity. We have grown also other QWs with In content 22-24% (results not shown) without a preQW layer and for all these samples the result was similar, i.e. very weak PL. Since the built-in fields in the 12-nm-wide QWs do not impact the luminescence intensity as much as in 2.6-nm-wide counterparts, we can conclude that preQW plays an important role in reduction of nonradiative recombination centers related to most likely point defects or other structural defects.

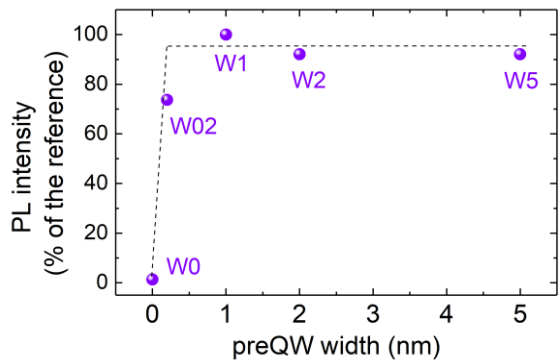

*Figure 4. Relative PL intensity of 12 nm thick QW structures as a function of preQW thickness. Dashed lined are shown to guide the eye.*

For staggered QWs we have indium content of 8% before the QW and 16% and 24% in the QW – therefore the In composition change is about 8 percentage points at each interface. For standard QWs the In content was 8% before QW and 24% in the QW - resulting in In content difference of 16 percentage points. Perhaps it is a key factor which can change growth mechanism of efficient QWs. Below we demonstrate that the density of nonradiative recombination centers and related other structural defects is greatly reduced for staggered QWs.

**Cathodoluminescence studies**

In order to get an insight into the origins of differences in PL intensity between standard QWs and their staggered counterparts, we investigated the thin double QW structures by cathodoluminescence imaging (CL) at room-temperature. Monochromatic maps taken at peak QW wavelength have been collected for samples A0, A1, B0 and B1. Peak wavelengths observed in CL differ slightly from the ones measured by PL due to different excitation powers. Figure 5(a,b) and Figure 5(c,d) compare the CL results for samples A0-A1 and B0-B1, respectively. Two of the dark spots present in Figure 5 (b), marked in circles, were also observed in CL map collected at λ=365nm (map is not presented here), which suggests that these defects were not generated during the growth of the QW but rather originate from the GaN substrate. Two conclusions can be drawn from the analysis of the CL maps shown in Figure 5. First, comparing the emission from the QWs between samples A0 and A1 we see that the number of dark spots and their density is much lower for the staggered QW (A1) presented in Figure 5(b) than for standard QW (A0) presented in Figure 5(a). The same observation holds for B0 and B1 samples, for which the peak emission mapping showed a more uniform emission from the staggered QW (B1) than from the standard counterpart (B0), cf. Figure 5(d) and Figure 5(c), respectively. Second, by comparison of sample A0 to B0 and A1 to B1, we see that the higher the In content of the QW, the higher the density of dark spots in CL maps, both for standard and staggered QWs. The CL results show clearly, that the decrease of PL for high In content QWs is related with the generation of defects. Reduction of these defects through insertion of the preQW increases the homogeneity of emission observed in CL and significantly increases the PL intensity.

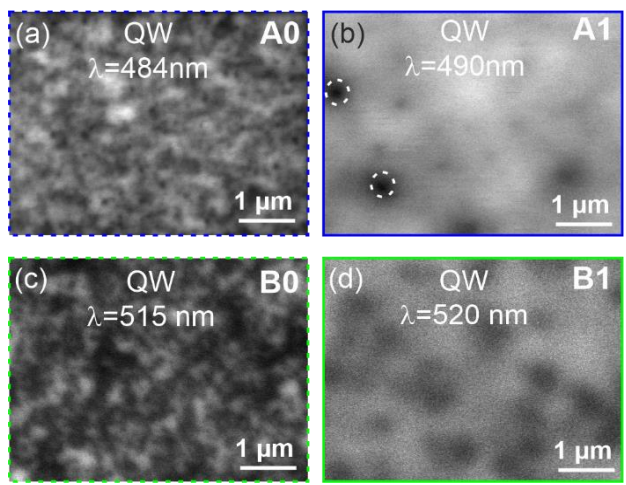

Figure 5. CL monochromatic maps of: (a) A0, (b) A1, (c) B0 and (d) B1 samples taken at the wavelength corresponding to the QW emission.

**Identification of defects by TEM**

Selected samples were studied by cross-sectional TEM in order to investigate the presence of extended defects: two samples without preQW – B0 and W0 and one sample with a 2-nm preQW – W2. The cross-section studies of samples B0 and W0, confirmed the presence of long basal-plane stacking faults (BSFs) in the QWs. The bright field (BF) STEM images presented in Figure 6(a,b) show numerous threading dislocations (TDs) starting from the QWs. The dark-field two beam investigation of such areas with TDs shows that the dislocations originate from BSF formed in the QW, cf. Figure 6(c). The observed TDs are geometrically necessary dislocations formed at the folds of prismatic stacking fault (PSF) domains terminating the BSFs. The mechanism of formation of TDs from BSF/PSF domains was described elsewhere [24, 25]. The I1 BSFs were repeatedly found to form at the bottom interface between the InGaN barrier and high-In content QW layer, as shown in high resolution TEM image shown on Figure 6(d). Except from the BSFs and TDs, we also found voids present at the bottom interface of the InGaN QW in the structure W0. The presence of voids is manifested by the local roundish features caused by the strain due to empty space in the crystal lattice, as marked with arrows in Figure 6(e,f) and additionally marked with a circle in Figure 6(f). Similar features were observed for thermally annealed InGaN QWs, in which structural degradation was initiated by formation of voids, caused by agglomeration of vacancies diffusing to the QWs during annealing at the temperatures on average 200°C higher that the growth temperature of the QW [26]. Presence of voids in the annealed InGaN QWs was accompanied by the formation of long I1 BSFs, pointing out to the same origin of both defects. The classic model for the formation of the I1 BSFs assumes removal of the one basal plane caused by the agglomeration of vacancies [18] and as well the only possible way of void formation in the studied QW structures is by agglomeration of vacancies. Since the studied structures were not annealed and the growth temperatures for MBE technique are too low in order to activate the diffusion of vacancies, it is more plausible that the agglomeration of the point defects takes place during the abrupt change of the In composition from 8% to 22% or from 8% to 24%. Note that all the observed defects: voids, BSFs and TDs originate at the bottom interface of the QW, where the composition difference between the layers is the highest. The last two TEM pictures presented in Figure 6(g,h) show HAADF STEM images of the structure W2. In this case no structural defects such as TDs, BSFs or voids are visible, which is consistent with the high PL intensity observed for this QW. In Figure 6(h) magnification of the wide QW of W2 sample shows that a 2-nm preQW was formed during epitaxy, as intended.

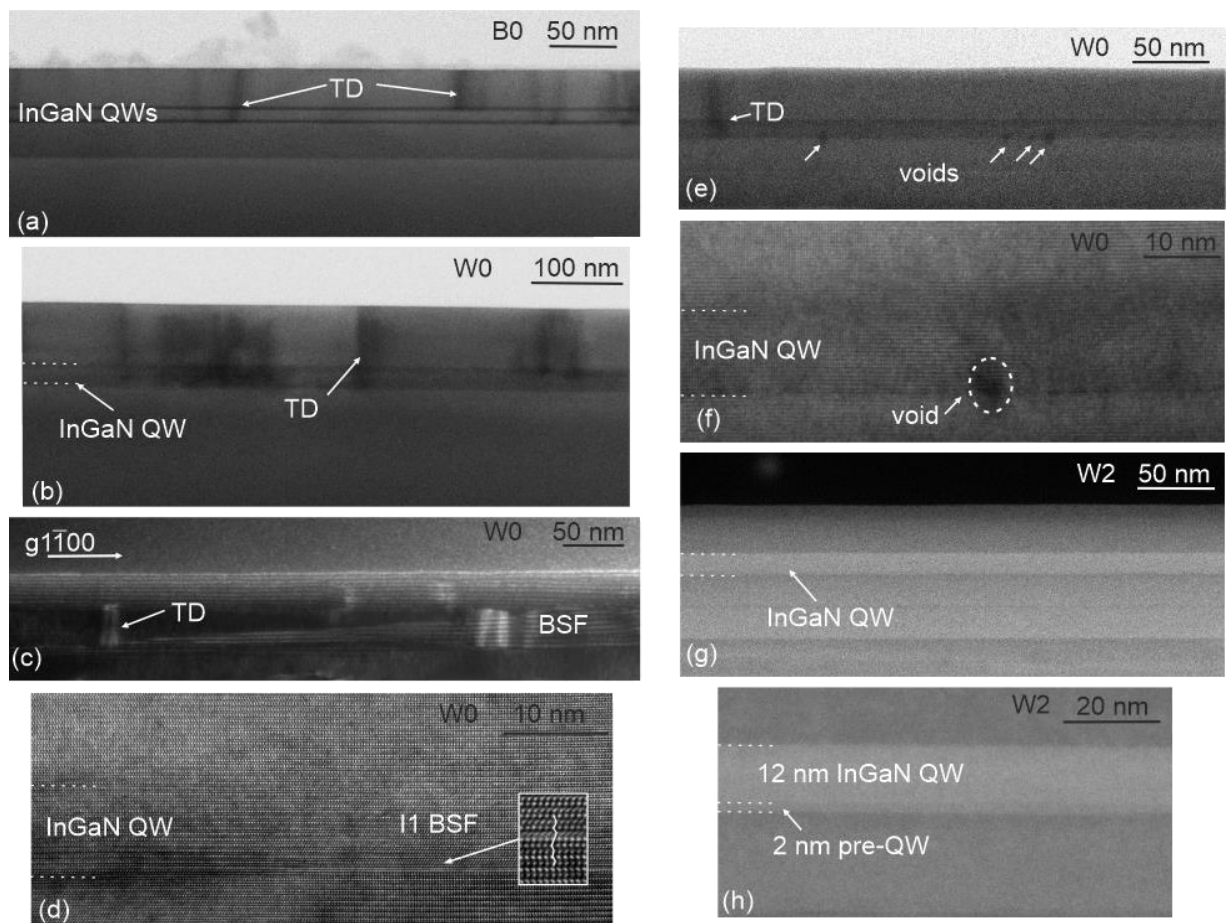

*Figure 6. Cross-section TEM images of samples B0 and W0 evidencing the presence of defects in the InGaN QWs. (a) and (b) Bright field (BF) STEM images of sample B0 and W0, respectively. Threading dislocations (TD) starting from the QWs are indicated; (c) Dark field TEM image of the W0 structure taken with g1-100 showing long BSF in the QW and TDs starting at the BSF; (d) HRTEM image taken along [11-20] zone axis of the W0 structure showing the I1 BSF at the bottom of the QW, the stacking*

*sequence of the basal 0002 planes across the BSF is indicated in the inset, (e) BF STEM image of the W0 sample showing voids present at the lower interface of the QW with the barrier; (f) HRTEM image of the W0 sample taken along [11-20] zone axis showing the roundish area of the lower intensity caused by strain due to the void present at the lower QW interface; (g) and (h) HAADF STEM images of the structure W2, no structural defects are observed in the field of view.*

**Discussion**

Higher quantum efficiency of the staggered QWs in reference to the standard, rectangular composition profile counterparts is usually explained by the reduction of the electric field. However, we observed in increase in PL intensity for staggered QWs even with a lower overlap of electron-hole wavefunctions. Therefore, there has to be additionally a strong Shockley-Read-Hall (SRH) nonradiative recombination channel present in standard QWs, such as structural defects. We experimentally demonstrated high density dark spots for QWs without preQW, which is an indication for defects responsible for SRH nonradiative recombination. For samples with preQW CL was very uniform. Confirming the presence of point defects in the QWs is very challenging and usually indirect. However, the formation of defects at the bottom interface of high-In-content QWs is corroborated by TEM where voids, BSFs and TDs are found only in the samples without preQW. The formation of such structural defects requires large number of vacancies. Therefore, we attribute the quenching of the PL to the presence of large number of vacancies in QW for the structures without preQW. Growth of the preQW of an intermediate indium content between the barrier and the well prevents generation of vacancies leading as a result to a large increase of the PL intensity.

**Conclusions**

We have shown that staggered QWs grown by PAMBE have higher luminescence intensity as compared to standard, rectangular composition profile QWs. In our work we studied standard and staggered construction for two thicknesses of QWs: 2.6 nm and 12 nm. The study provides new evidences on the consequences of the application of thin preQW on the luminescence and structural properties of QW structures. We demonstrate an additional unexpected benefit of staggered QWs, namely the improved optical quality that is not solely a result of higher wavefunction overlap. When a preQW layer of an intermediate In composition is inserted below the QW, point defects formation is reduced and therefore the optical quality of the staggered QW is higher. Standard QWs CL show a high dark spots density. TEM studies evidence presence of voids, BSFs and TDs at the bottom interface of the QW that are believed to be formed through aggregation of point defects. The presented CL and TEM results support the hypothesis that the use of preQW of an intermediate In content leads to reduction of point defects density, which in turn are responsible for increase of nonradiative recombination rate. Lower number of nonradiative recombination centers explains the increased quantum efficiency reported for staggered QWs.

**Acknowledgements**

Authors would like to thank Grzegorz Staszczak for CL and Marcin Kryśko for XRD measurements.

**Funding**

This work received funding from the Foundation for Polish Science co-financed by the European Union under the European Regional Development Fund within TEAM-TECH POIR.04.04.00-00-210C/16-00 and POWROTY/REINTEGRATION POIR.04.04.00-00-4463/17-00 projects. This work was also financially supported by National Science Centre Poland within grants SONATA no. 2019/35/D/ST5/02950 and 2019/35/D/ST3/03008. The research leading to these results has also received funding from the Norway Grants 2014-2021 via the National Centre for Research and Development within grant no. NOR/SGS/BANANO/0164/2020 and from National Centre for Research and Development grant no. LIDER/29/0185/L-7/15/NCBR/2016, LIDER/35/0127/L9/17/NCBR/2018 and no. LIDER/287/L-6/14/NCBR/2015.